# Frequentist performances of Bayesian prediction intervals for random-effects meta-analysis

Yuta Hamaguchi
Department of Statistical Science, School of Multidisciplinary Sciences, The Graduate University for Advanced Studies, Tokyo, Japan
Diagnostics Department, Asahi Kasei Pharma Corporation, Tokyo, Japan

Hisashi Noma, PhD*
Department of Data Science, The Institute of Statistical Mathematics, Tokyo, Japan
ORCID: http://orcid.org/0000-0002-2520-9949

Kengo Nagashima, PhD
Research Center for Medical and Health Data Science, The Institute of Statistical Mathematics, Tokyo, Japan

Tomohide Yamada, MD, PhD
Department of Diabetes and Metabolic Diseases, Graduate School of Medicine, The University of Tokyo, Tokyo, Japan

Toshi A. Furukawa, MD, PhD
Departments of Health Promotion and Human Behavior, Kyoto University Graduate School of Medicine/School of Public Health, Kyoto, Japan

*Corresponding author: Hisashi Noma
 Department of Data Science, The Institute of Statistical Mathematics
 10-3 Midori-cho, Tachikawa, Tokyo 190-8562, Japan
 TEL: +81-50-5533-8440
 e-mail: noma@ism.ac.jp

**Summary**

The prediction interval has been increasingly used in meta-analyses as a useful measure for assessing the magnitude of treatment effect and between-studies heterogeneity. In calculations of the prediction interval, although the Higgins–Thompson–Spiegelhalter method is used most often in practice, it might not have adequate coverage probability for the true treatment effect of a future study under realistic situations. An effective alternative candidate is the Bayesian prediction interval, which has also been widely used in general prediction problems. However, these prediction intervals are constructed based on the Bayesian philosophy, and their frequentist validities are only justified by large-sample approximations even if non-informative priors are adopted. There has been no certain evidence that evaluated their frequentist performances under realistic situations of meta-analyses. In this study, we conducted extensive simulation studies to assess the frequentist coverage performances of Bayesian prediction intervals with 11 non-informative prior distributions under general meta-analysis settings. Through these simulation studies, we found that frequentist coverage performances strongly depended on what prior distributions were adopted. In addition, when the number of studies was smaller than 10, there were no prior distributions that retained accurate frequentist coverage properties. We also illustrated these methods via applications to eight real meta-analysis datasets. The resultant prediction intervals also differed according to the adopted prior distributions. Inaccurate prediction intervals may provide invalid evidence and misleading conclusions. Thus, if frequentist accuracy is required, Bayesian prediction intervals should be used cautiously in practice.

# 1. Introduction

Random-effects models have been the primary statistical tools for meta-analyses since they enable quantitative evaluation of the average treatment effects accounting for the between-studies heterogeneity [1,2]. Conventionally, the "grand mean" parameter has been addressed as a primary estimand of these evidence synthesis studies, but its ability to express a summarized measure of treatment effects among multiple studies is substantially limited because it is solely a point measure that corresponds to the mean of the random-effects distribution. To quantify the true effect and effectiveness in real-world uses of the treatment, more appropriate measures are required that suitably reflect the degree of heterogeneity and magnitude of the treatment effect. The prediction interval was proposed to address this problem; it is defined as an interval that covers the true treatment effect in a future study with certain probability [3,4].

The prediction interval has been gaining prominence in recent meta-analyses because it enables the simultaneous assessment of uncertainties in treatment effects and heterogeneity between studies [5,6]. The Higgins–Thompson–Spiegelhalter (HTS) method [3] has been most widely used for calculation of the prediction interval. The HTS method is computational efficient and practically useful, but in a recent study by Partlett and Riley [7], the HTS prediction interval can have poor coverage properties when the number of studies $n$ is small. The most important reason for this is that HTS method is based on the large-sample approximation in which $n$ is sufficiently large. In medical meta-analyses, $n$ is usually less than 20 [8]; therefore, the large-sample approximation can be violated, similar to the problem with inference of the grand mean [9-11].

An effective alternative approach is Bayesian prediction methods, which have been extensively investigated for general statistical prediction problems [12,13]. Higgins et al. [3] also discussed Bayesian approaches in the context of constructing the prediction interval for meta-analyses. Although the Bayesian approach is always "exact" if it is used as a purely subjective method, most cases adopt non-informative prior distributions and use as objective Bayesian approaches [14,15]. When using non-informative prior distributions, the resultant predictions (inferences) are good approximations of frequentist predictions (inferences) [12,13,16]; accordingly, these methods have often been used in practice as if they were frequentist methods. However, a Bayesian prediction interval is rigorously defined as an interval within which a future value falls with a particular subjective probability, and its concordance with the frequentist probability is only assured under large-sample settings [12]. In practical situations of meta-analyses in medical research, $n$ is limited (usually < 20) [8]. Also, several simulation-based studies have indicated that Bayesian inferences are not necessarily accurate in the frequentist sense [17], and their frequentist performances vary dramatically among prior distributions [18]. To date, there is no definitive evidence regarding how Bayesian prediction intervals perform in practical meta-analysis situations.

In this article, we conducted extensive simulation studies to assess the frequentist coverage performances of the Bayesian prediction intervals with various 11 non-informative priors. In addition, we provide eight illustrative examples of analyses selected from recently published systematic reviews in leading medical journals (*BMJ*, *Lancet*, and *JAMA*). Our aim was to obtain certain numerical evidence for the operating characteristics of Bayesian prediction methods, as well as to provide recommendations for the practical use of these methods. We also provide related simulation and real data analysis results for the credible intervals in the Supporting Information.

## 2. Prediction intervals for meta-analysis

We consider that there are $n$ clinical trials and $y_i$ $(i = 1,2,...,n)$ is the estimated treatment effect measure in the $i$th trial. Commonly used effect measures are mean difference, standardized mean difference, risk difference, risk ratio, and odds ratio [19,20]; the ratio measures are typically transformed to logarithmic scales. The random-effects model in meta-analyses [1,2] is then defined as

$$y_i \sim N(\theta_i, \sigma_i^2) \tag{1}$$
$$\theta_i \sim N(\mu, \tau^2)$$

where $\theta_i$ is the true effect size of the $i$th study, and $\mu$ is the grand mean parameter. $\sigma_i^2$ and $\tau^2$ express within- and across-studies variances; $\sigma_i^2$ is usually assumed to be known and fixed to their valid estimates. The across-studies variance $\tau^2$ represents the degree of heterogeneity across studies. Conventionally, the grand mean parameter $\mu$ is used as a summary measure of the random-effects meta-analysis as an average treatment effect, and it is estimated as $\hat{\mu} = (\sum_{i=1}^{n} \hat{w}_i y_i)/(\sum_{i=1}^{n} \hat{w}_i)$, where $\hat{w}_i = (\sigma_i^2 + \hat{\tau}^2)^{-1}$; $\hat{\tau}^2$ is an estimator of the heterogeneity variance, e.g., the method of moment estimator proposed by DerSimonian and Laird [1]. However, when certain heterogeneity exists, the point measure has a limited ability to utilize meta-analysis evidence for medical policy making and health technology assessments; its effectiveness should be evaluated considering the heterogeneity in the target population. Thus, the prediction interval has been gaining prominence as a way to add useful information involving the uncertainty of the treatment effect and its heterogeneity [5,6].

The $100(1 − \alpha)\%$ prediction interval in meta-analysis is formally defined as an interval that covers the treatment effect $\theta_{new}$ in a future study with $100(1 − \alpha)\%$ probability [3,4]. Higgins et al. [3] proposed a simple plug-in type prediction interval,

$$\left[\hat{\mu} - t^{\alpha}_{n-2}\sqrt{\hat{\tau}^2_{DL} + \widehat{Var}[\hat{\mu}]}, \hat{\mu} + t^{\alpha}_{n-2}\sqrt{\hat{\tau}^2_{DL} + \widehat{Var}[\hat{\mu}]}\right]$$

where $\hat{\tau}^2_{DL}$ is the DerSimonian-Laird's method-of-moment estimator [1] of $\tau^2$, $\widehat{Var}[\hat{\mu}] = 1/(\sum_{i=1}^{n}(\sigma_i^2 + \hat{\tau}^2_{DL})^{-1})$ is the variance estimator of $\hat{\mu}$, and $t^{\alpha}_{n-2}$ is the $100(1 - \alpha/2)$ percentile of the *t*-distribution with $n - 2$ degrees of freedom. The HTS prediction interval is based on two approximations: $(\hat{\mu} - \mu)/\sqrt{\widehat{Var}[\hat{\mu}]}$ is approximately distributed as N(0, 1), and $(n - 2)(\hat{\tau}^2_{DL} + \widehat{Var}[\hat{\mu}])/(\tau^2 + \widehat{Var}[\hat{\mu}])$ is approximately distributed as $\chi^2(n - 2)$. These approximations are generally not accurate under small- or moderate-$n$ settings. Thus, in the simulation studies of Partlett and Riley [7] and Nagashima et al. [21], the coverage probability of the prediction interval for $\theta_{new}$ was below the nominal level under general settings of meta-analyses of medical studies, especially when $n < 20$ [8]. To address the undercoverage problems, Partlett and Riley [7] proposed other plug-in–type intervals instead of $\hat{\tau}^2_{DL}$, e.g., the restricted maximum likelihood (REML) estimator or the Hartung–Knapp estimator [22]. Also, Nagashima et al. [21] proposed a parametric bootstrap–based approach using a confidence distribution. In their extensive simulation studies, Nagashima et al. [21] showed that the coverage probability of their prediction interval accorded to the nominal level consistently. These prediction intervals are all calculable using the `pimeta` package [23] in R (R Foundation for Statistical Computing, Vienna, Austria).

## 3. Bayesian prediction intervals

### 3.1 The Bayesian hierarchical model and prediction

The Bayesian framework using the Markov Chain Monte Carlo (MCMC) is an established prediction method in statistics, and represents an alternative effective approach for the prediction problem. Higgins et al. [3] also discussed the uses of the Bayesian methods in addressing the prediction problem in meta-analyses. In the Bayesian framework, a prediction distribution for the effect $\theta_{new}$ is computable by sampling a new study, $\theta_{new} \sim N(\mu, \tau^2)$ [3,24] by MCMC. A 95% prediction interval for the new study is obtained simply from the 2.5% and 97.5% quantiles of the posterior distribution of $\theta_{new}$.

A Bayesian hierarchical model for the random-effects model (1) is ordinarily constructed assuming prior distributions for the parameters of random-effects distribution $\mu$ and $\tau^2$. The purely Bayesian predictions can be conducted under this framework, but here we consider using this Bayesian framework to conduct approximate frequentist predictions using non-informative priors, i.e., to provide an accurate prediction interval with a sense of frequentist probability.

*3.2 Prior distributions for approximate frequentist prediction*

Various non-informative prior distributions are considered for the meta-analysis model [18]. Lambert et al. [18] also conducted large simulation studies to assess the frequentist performances of the inferences of $\mu$ and $\tau^2$, and concluded that they strongly depended on the prior distribution of $\tau^2$. Following their discussions, we considered 11 non-informative priors involving improper priors to assess the impact on prediction accuracy, which is involved in the recently developed bayesmeta package in Röver [15], which uses various modern methods to select a prior distribution. For the grand mean parameter $\mu$, we consistently used a diffuse Gaussian distribution,

$$\mu \sim N(0, 10000)$$

Also, we considered the standard factorable prior distribution, which can be factored into independent marginal, $p(\mu, \tau) = p(\mu) \times p(\tau)$. The 11 prior distributions for $\tau^2$ considered here were as follows.

### 3.2.1 Uniform prior distributions for $\tau$

$$p(\tau) \propto 1$$

This prior distribution is the most intuitive flat non-informative improper prior for the heterogeneity standard deviation parameter $\tau$ [15]. In addition, as a popular choice in these Bayesian analyses, a proper uniform prior on a limited range can be also considered. One choice would be $\tau \sim \text{Uniform}(0,10)$, which is a flat uniform distribution on (0, 10).

### 3.2.2 Uniform prior distribution in $\sqrt{\tau}$

$$p(\tau) \propto \frac{1}{\sqrt{\tau}}$$

This is the uniform prior in $\sqrt{\tau}$. It has been proposed that a requirement is reasonable for non-informative priors, because it has invariance with respect to re-scaling of $\tau$ [25,26]. Due to this requirement, a family of improper prior distribution with density is $p(\tau) \propto \tau^a$ ($-\infty < a < \infty$). As a special case, this prior distribution corresponds to $a = -0.50$ expressing a monotonically decreasing density function; $a = 0$ corresponds to the improper uniform prior in Section 3.2.1.

### 3.2.3 Jeffreys prior distribution

$$p(\tau) \propto \sqrt{\sum_{i=1}^{n} \left(\frac{\tau}{\sigma_i^2 + \tau^2}\right)^2}$$

This is the non-informative Jeffreys prior [27], which is formally given by $p(\mu,\tau) \propto \sqrt{\det(J(\mu,\tau))}$ where $J(\mu,\tau)$ is the Fisher information matrix. In the present Bayesian hierarchical model, the two parameters $\mu$ and $\tau^2$ are orthogonal in the sense that the off-diagonal elements of the Fisher information matrix are 0, and the Jeffreys prior corresponds to the Tibshirani's non-informative prior [28]; then, the prior density function is given above. This prior also corresponds to the Berger–Bernardo reference prior [14,29].

*3.2.4 Berger–Deely prior distribution*

$$p(\tau) \propto \prod_{i=1}^{n}\left(\frac{\tau}{\sigma_i^2 + \tau^2}\right)^{1/n}$$

This is another variation of the Jeffreys prior, provided by Berger and Deely [30]. This prior distribution is also improper, and is concordant with the Jeffreys prior when all within-study variances $\sigma_i^2$ are equal.

*3.2.5 The proper conventional prior distribution*

$$p(\tau) \propto \prod_{i=1}^{n}\left(\frac{\tau}{(\sigma_i^2 + \tau^2)^{3/2}}\right)^{1/n}$$

This is a proper variation of the Jeffreys prior that was proposed by Berger and Deely [30]. This prior distribution is intended as a non-informative, but is used as a proper one for testing or model selection purposes [30] [31].

*3.2.6 DuMouchel prior distribution*

$$p(\tau) = \frac{s_0}{(s_0 + \tau)^2}, \quad s_0^2 = \frac{n}{\sum_{i=1}^{n} \sigma_i^{-2}}$$

This is the DuMouchel prior distribution [16,32] for the heterogeneity parameter $\tau$. $s_0^2$ is the harmonic mean of within-study variances $\sigma_i^2$. This prior corresponds to a log-logistic distribution for $\tau$ that has the mode at 0 and the median at $s_0$.

### 3.2.7 Uniform shrinkage prior distribution

$$p(\tau) = \frac{2s_0^2 \tau}{(s_0^2 + \tau^2)^2}, \quad s_0^2 = \frac{n}{\sum_{i=1}^{n} \sigma_i^{-2}}$$

This prior distribution is derived as a uniform prior for the average shrinkage factor $S_0(\tau) = s_0^2/(s_0^2 + \tau^2)$ [16,33]. The median is also $s_0$, and the forms of the DuMouchel prior and this prior depend on the harmonic mean $s_0$. The uniform prior for $S_0(\tau)$ is equivalent to a uniform prior of $1 - S_0(\tau) = \tau^2/(s_0^2 + \tau^2)$, which has similar form to the Higgins' $I^2$ statistic [34].

### 3.2.8 Uniform prior distribution in $I^2$ statistic

$$p(\tau) = \frac{2\hat{\sigma}^2 \tau}{(\hat{\sigma}^2 + \tau^2)^2}, \quad \hat{\sigma}^2 = \frac{(n-1)\sum_{i=1}^{n} \sigma_i^{-2}}{(\sum_{i=1}^{n} \sigma_i^{-2})^2 - \sum_{i=1}^{n} \sigma_i^{-4}}$$

This is the uniform prior distribution for Higgins' $I^2$ statistic [34]. As mentioned in Section 3.2.7, this prior density is obtained by substituting the harmonic mean $s_0^2$ for their average $\hat{\sigma}^2$ from the uniform shrinkage prior; these two priors have similar forms.

### 3.2.9 Proper inverse-Gamma prior distributions

$$\frac{1}{\tau^2} \sim \text{Gamma}(0.001, 0.001)$$

This prior distribution is the most commonly used semi-conjugate prior for the heterogeneity variance parameter [18]. The shape of this prior distribution is mostly flat over a wide range, but has a small mode near 0. In addition, we considered a variation of this prior,

$$\frac{1}{\tau^2} \sim \text{Gamma}(0.01, 0.01)$$

whose two parameters were changed. This prior is also a vague prior for $\tau^2$, but more informative than the above prior. Using this prior, we can assess the sensitivity of altering the hyperparameters to the frequentist performance of the prediction interval.

## 4. Simulations

### *4.1 Designs and settings*

We conducted a series of simulation studies to provide certain evidence for frequentist performances of Bayesian prediction intervals for meta-analyses. We adopted 11 priors for $\tau^2$ explained in Section 3, (1) the uniform improper prior in $\tau$ (**Uniform**), (2) the uniform improper prior in $\sqrt{\tau}$ (**Sqrt**), (3) the Jeffreys prior (**Jeffreys**), (4) the Berger–Deely prior (**Berger–Deely**), (5) the proper conventional prior (**Conventional**), (6) the DuMouchel prior (**DuMouchel**), (7) the uniform shrinkage prior (**Shrinkage**), (8) the uniform prior for $I^2$ statistic (**I2**), (9) the proper uniform prior $U(0, 10)$ (**Proper 1**), (10) the proper inverse-Gamma prior $\text{Gamma}(0.001, 0.001)$ (**Proper 2**), and (11) the inverse-Gamma prior $\text{Gamma}(0.01, 0.01)$ (**Proper 3**). The 95% prediction intervals for a future study were calculated based on the 2.5% and 97.5% quantiles of the posterior distribution of $\theta_{new} \sim N(\mu, \tau^2)$. We used the `bayesmeta` package [15] in R and OpenBUGS [35] for computations. Also, we added the HTS interval (**HTS**) as a reference method.

The simulation data were generated mimicking the simulation settings of Brockwell and Gordon [9,36], which consider typical settings of meta-analyses in medical studies that assess an overall odds-ratio. The grand mean parameter $\mu$ was set to 0, without loss of generality for assessing coverage and precision of the prediction intervals. The within-study variances $\sigma_i^2$ were generated from a chi-squared distribution with 1 degree of freedom, multiplied by 0.25, and truncated within an interval [0.009, 0.6]. The number of studies $n$ and the heterogeneity variance $\tau^2$ were varied for two patterns: (1) $n$ was fixed to 7 or 15, and $\tau^2$ varied among 0.01, 0.02,…,0.20, and (2) $\tau^2$ was fixed to 0.10 or 0.20, and $n$ varied among 4, 5, 6,…,20. For each scenario, we simulated 10000 replications. We assessed the empirical covariate rates of randomly generated $\theta_{new} \sim N(\mu, \tau^2)$ of the 12 prediction intervals and their empirical expected widths for the 10000 results. The coverage probabilities are desirable to accord the nominal level, 95%. We also assessed the coverage performances of credible intervals for the 11 priors. The results are presented in e-Appendix A in Supporting Information.

*4.2 Results*

First, the results of the simulation studies for the scenarios in which $n$ was fixed and $\tau^2$ was varied are presented in Figure 1. Under both settings for the number of studies ($n = 7, 15$) settings, most of the Bayesian methods exhibited overcoverage when $\tau^2 = 0.01$; only HTS had coverage probabilities around 0.95. However, when $\tau^2$ grew larger, HTS exhibited undercoverage. HTS consistently exhibited undercoverage when $\tau^2 > 0.01$, and these results were consistent with previous simulation studies [7,21]. Under $n = 7$, the coverage probabilities for Sqrt, DuMouchel, and Proper 2 were below the nominal level when there was certain heterogeneity. Under $n = 15$, they also exhibited undercoverage under smaller $\tau^2$. However, their coverage probabilities were generally larger when $\tau^2$ grew larger. The simulation results of Shrinkage and I2 had similar trends because of the similarities in the shapes of their prior distributions. These two methods also exhibited overcoverage when $\tau^2 < 0.05$; but as heterogeneity got larger, they tended to exhibit undercoverage. For Uniform, Jeffreys, Berger–Deely, Conventional, Proper 1, and Proper 3, the coverage probabilities were above the nominal level regardless of the degree of heterogeneity under $n = 7$; when $\tau^2 = 0.20$, the coverage probabilities of Conventional and Proper 3 were around 0.95. Also, their expected widths were larger than those of the aforementioned priors. Under $n = 15$, they exhibited overcoverage when $\tau^2 < 0.10$, and had accurate coverage probability around the nominal level when $\tau^2 \geq 0.10$, except for Proper 3. Proper 3 exhibited undercoverage when $\tau^2 \geq 0.10$; also, only Berger–Deely tended to exhibit overcoverage. The results of Proper 2 and Proper 3 were quite different, although they were the same parametric inverse-gamma priors; the differences indicate sensitivity to the selection of hyperparameters, and suggest that these trends might change if the simulation settings are altered. These results reveal that we cannot explicitly specify which priors can provide accurate prediction intervals in general. The expected widths clearly reflected the trends of coverage probabilities. From the above, Uniform, Jeffreys, Berger–Deely, Conventional, Proper 1 had accurate frequentist coverage properties when $n = 15, \tau^2 \geq 0.10$. However, the others exhibited undercoverage or overcoverage, and did not provide accurate prediction intervals in the frequentist sense.

Second, the results of simulation studies for the scenarios in which $\tau^2$ was fixed and $n$ was varied are presented in Figures 2. Under $\tau^2 = 0.10$, when the number of studies were extremely small ($n = 4,5$), DuMouchel, Shrinkage, I2, Proper 2, and HTS had the most accurate coverage probabilities. However, as $n$ got larger, these five methods exhibited undercoverage. In particular, the coverage probabilities of DuMouchel were below 0.90. For Sqrt, the coverage probabilities were above the nominal level when $n \leq 6$, but far below 0.95 when $n > 6$. Uniform, Jeffreys, Berger–Deely, Conventional, Proper 1, and Proper 3 tended to exhibit overcoverage when $n < 12$; when $n \geq 12$, they had accurate coverage probabilities except for Proper 3, which exhibited undercoverage then. Among these priors, Berger–Deely exhibited the greatest degree of overcoverage. Under $\tau^2 = 0.20$, DuMouchel, Shrinkage, I2, Proper 2 and HTS consistently exhibited undercoverage, but when $n$ grew larger, their coverage performances improved. In addition, the coverage probabilities of Sqrt and Conventional were above the nominal level under $n \leq 5$ or $6$, but they tended to exhibit undercoverage when $n > 6$, with Sqrt exhibiting undercoverage to a greater degree. Uniform, Jeffreys, Berger–Deely, Proper 1 exhibited overcoverage under $n \leq 10$, but they had generally accurate coverage probabilities when $n > 10$. Proper 3 exhibited overcoverage when $n \leq 6$, but exhibited undercoverage when $n$ grew larger. In addition, under these settings, the coverage performances of Proper 2 and Proper 3 were also quite different. When $n$ grew larger, the differences got to be smaller, because the relative information of observed data got larger. The expected widths also clearly reflected the trends in coverage probabilities; Uniform and Proper 1 had markedly larger expected widths than the other priors when $n = 4$ or $5$. Overall, Uniform, Jeffreys, Berger-Deely, Proper 1 had accurate frequentist coverage properties when $\tau^2 = 0.10, n \geq 12$ and $\tau^2 = 0.20, n \geq 11$; Conventional also achieved nearly accurate coverage performance. However, the others exhibited undercoverage or overcoverage, and did not provide accurate prediction intervals in the frequentist sense. Therefore, there are not favourable prediction intervals, especially under $n < 10$, that we can recommend as accurate prediction tools in frequentist sense in meta-analyses of medical studies.

## 5. Real data examples

For illustrative purposes, we applied Bayesian prediction intervals to eight real data examples that were recently published in leading medical journals (*BMJ*, *Lancet* and *JAMA*). The summary of eight datasets is presented in Table 1. To describe operational characteristics of various prediction intervals in details, we chose the example datasets with various characteristics: the numbers of studies were distributed among 3 to 22, and their heterogeneity variance estimates were valued from 0.00 to 0.60. We present the Bayesian prediction intervals using the 11 reference priors adopted in Section 4. In addition, as reference methods, we also present the HTS interval (**HTS**), the HTS interval using the Hartung–Knapp variance estimator (**HTS–HK**), the HTS interval using the Sidik–Jonkman bias-corrected variance estimator (**HTS–SJ**) [7], and the prediction interval–based parametric bootstrap approach using the confidence distribution (**pimeta**) of Nagashima et al. [21] as reference methods.

The results are presented in Figures 3 and 4. We also present the 95% credible intervals for the grand mean parameter $\mu$ for the eight meta-analyses. The results are presented in e-Appendix B in Supporting Information.

### 5.1 CPR data

Hüpfl et al. [37] conducted a meta-analysis to assess the association of chest-compression–only cardiopulmonary resuscitation (CPR) with survival in patients who experience out-of-hospital cardiac arrest. The outcome was survival to hospital discharge or 30-day survival, and the outcome measure was the risk ratio (RR). Figure 3 (a) presents the 95% prediction intervals for this dataset. This meta-analysis included three studies. The DerSimonian-Laird heterogeneity variance estimate was $0.00$. For this dataset, the Uniform and Proper 1 prediction intervals were the widest, and the widths were several times greater than the narrowest (DuMouchel and Shrinkage). These results were consistent with simulation results in which the expected widths of Uniform and Proper 1 were especially large when $n$ was small. Also, the widths of the prediction intervals were quite different among different prior distributions. Berger–Deely, Jeffreys, Sqrt, and Proper 3 also provided relatively wide prediction intervals. Compared with the pimeta and HTS methods, these results might lead to different interpretations. However, all prediction intervals included 1, potentially indicating that there were some subpopulations in which the treatment effect was null or harmful; although the overall RR was 1.215 (95%CI: 1.009, 1.464).

*5.2 Corticosteroids data*

This meta-analysis was conducted to estimate the benefits and harms of using corticosteroids as an adjunct treatment for sore throat [38]. The outcome was complete resolution of pain and the effect measure was RR. Figure 3 (b) presents the 95% prediction intervals of this dataset. The data included five studies and indicated a certain heterogeneity ($\hat{\tau}^2_{DL} = 0.33, p = 0.014$). The Uniform and Proper 1 intervals were also the widest among the comparators. Besides, the Shrinkage and I2 intervals were the narrowest. Among the frequentist methods, the HTS, HTS–HK, HTS–SJ, and pimeta intervals were not so different. All prediction intervals were quite wide and included 1, and they might indicate there are some subpopulations for which the treatment effect is null or harmful; although the overall RR was 2.233 (95% CI: 1.177, 4.235).

*5.3 QOL data*

Gaertner et al. [39] conducted a meta-analysis to assess the effect of specialist palliative care on quality of life (QOL). The outcome measures were different among the seven studies (e.g., European Organization for Research and Treatment of Cancer quality of life questionnaire; EORTC QLQ-C30), and were synthesized as standardized mean difference (SMD). Figure 3 (c) presents the 95% prediction intervals of this dataset. The heterogeneity variance estimate was the largest among the eight examples ($\hat{\tau}^2_{DL} = 0.60, p < 0.001$). The Uniform and Proper 1 intervals were also the widest, as in the former two examples. The I2 and Shrinkage intervals were the narrowest among the Bayesian prediction intervals, and HTS was the narrowest among all comparators. Also, the Berger–Deely interval was concordant with the Jeffreys interval, because the within-study variances for individual studies were equal. For this dataset, the overall SMD was 0.569 (−0.017, 1.154) and was not significant. Also, all prediction intervals included 0, suggesting that there were some subpopulations in which the treatment effect was null or harmful.

*5.4 PPI data*

Crocker et al. [40] investigated the impact of patient and public involvement (PPI) on rates of enrolment and retention by a systematic review of clinical trials. The outcome was patient enrolment rate, and the outcome measure was odds ratio (OR). Figure 3 (d) presents the 95% prediction intervals of this data. The data included eight studies and indicated small heterogeneity ($\hat{\tau}_{DL}^2 = 0.00$). The Berger–Deely interval was the widest for this dataset, and Conventional, Jeffreys, Uniform, Proper 1, and Proper 3 intervals were comparable with it. Although all the 11 Bayesian prediction intervals and pimeta interval included 1, the HTS, HTS–HK and HTS–SJ intervals did not. The HTS intervals indicated certain heterogeneity, but implied that the directions of OR are consistent in the target population. Besides, the other Bayesian prediction and pimeta methods indicated that the directions might change for certain subpopulations. The interpretations and conclusions of this meta-analysis might change among the different prediction methods.

*5.5 SBP data*

This dataset was provided as a hypothetical meta-analysis analysed in Riley et al. [4]. They generated a hypothetical dataset examining an effect of an antihypertensive drug. The outcome was systolic blood pressure (SBP; mmHg), and the outcome measure was SMD.

The dataset included 10 studies and exhibited moderate heterogeneity ($\hat{\tau}^2_{DL} = 0.03, p < 0.001$). Figure 4 (e) presents the 95% prediction intervals. The Uniform and Proper 1 intervals were the widest, and the Shrinkage interval the narrowest, in terms of Bayesian prediction intervals. The HTS–HK and HTS–SJ intervals were markedly wider than the HTS interval, because the heterogeneity variance estimate by DerSimonian–Laird's method was smaller than that of the REML method. Although there was some difference in prediction intervals, all prediction intervals included 0; the overall SMD was −0.334 (95%CI: −0.484, −0.184).

*5.6 DPP-4 data*

This meta-analysis was conducted to quantify the risk of hypoglycaemia with dipeptidyl peptidase-4 (DPP-4) inhibitors and sulphonylureas compared with placebo and sulphonylureas [41]. The outcome was an incidence of hypoglycaemia, and the effect measure was RR. Figure 4 (f) presents the prediction intervals of this data. The data included 10 studies and revealed moderate heterogeneity ($\tau^2_{DL} = 0.02$). The Berger–Deely, Conventional, Jeffreys, Uniform, Proper 1 and Proper 3 intervals were wider than the others. Also, HTS–HK and HTS–SJ were narrower than the HTS interval. Although the overall RR was 1.513 (95%CI: 1.219, 1.878), these prediction intervals might lead different interpretations of the results. The HTS–HK and HTS–SJ intervals did not include 1, and the Bayesian, HTS and pimeta intervals involved 1.

*5.7 Breakfast data*

Sievert et al. [42] examined the effect of regular breakfast consumption on weight change. The outcome was weight change (kg), and the outcome measure was mean difference (MD). Figure 4 (g) presents the 95% prediction intervals. This meta-analysis included 10 studies, and the heterogeneity variance estimate was 0.13. The Berger–Deely, Uniform, and Proper 1 intervals were wider than the others, and the DuMouchel interval was the narrowest. This trend was consistent with the simulation results in the conditions $n = 10$ and $\tau^2 = 0.1$ in Figure 2. There was a minimal difference in the prediction intervals, and all prediction intervals included 0. The HTS, HTS–HK, and HTS–SJ intervals were nearly equal, and slightly narrower than the pimeta interval. Although the grand mean was significantly larger than 0 (overall MD: 0.440, 95%CI: 0.065, 0.816), the direction of the effect might be changed for certain subpopulations.

*5.8 Pain data*

Häuser et al. [43] conducted a systematic review to compare the treatment effect of antidepressants on pain in patients with fibromyalgia syndrome. The outcomes were the pain questionnaire scores by VAS (visual analog scale), VASFIQ (visual analog scale fibromyalgia impact questionnaire) and NRS (numeric rating scale), and were synthesized as SMD. Figure 4 (h) presents the prediction intervals. The data included 22 studies, and the heterogeneity variance estimate was 0.03 (*p*=0.012). Among the Bayesian prediction intervals, the Berger–Deely, Conventional, Jeffreys, Uniform, Proper 1 and Proper 3 intervals included 0; on the other hand, the DuMouchel, I2, Shrinkage, Sqrt, and Proper 2 intervals did not. As for the frequentist methods, the HTS, HTS–HK, and HTS–SJ intervals did not include 0, but the pimeta interval did. Although the grand mean was significantly smaller than 0 (overall SMD: −0.427, 95%CI: −0.553, −0.302), the effect might be null for certain subpopulations. However, the interpretations and conclusions might differ based on the priors we adopted for this meta-analysis.

**6. Discussion**

The prediction interval has been gaining prominence in recent meta-analyses, and could become a standard statistical output in meta-analyses in the near future because of its effectiveness for the assessment of heterogeneity and uncertainties of treatment effects in target populations [3-6]. Bayesian prediction methods represent a useful approach in practices, but our study revealed that Bayesian prediction intervals are not necessarily accurate in the frequentist sense under practical situations.

Especially, when the number of studies $n$ is smaller than 10, our simulation studies showed that the Bayesian prediction intervals based on all 11 reference priors did not have favorable coverage performances. Also, the coverage performances were quite different. Thus, we must choose the prior distributions carefully. In particular, some priors exhibited serious undercoverage properties (Sqrt, DuMouchel, Shrinkage, I2, Proper 2, and Proper 3), and might under-estimate the heterogeneity and uncertainty in practice. Besides, Uniform, Jeffreys, Berger–Deely, Conventional, Proper 1 exhibited good coverage performances, in particular when $n > 10$; however, they tended to provide overly wide prediction intervals when $n \leq 10$, and might yield vague evidence.

If prediction intervals are too vague or too narrow, they can directly influence the conclusions of systematic reviews, providing misleading evidence for health technology assessments and policy making. Therefore, Bayesian prediction intervals should be carefully used in practice, and if there exist other accurate alternatives, they should be recommended. Certainly, it is not problematic if researchers wish to conduct purely Bayesian prediction with subjective probability, but most Bayesian applications in meta-analyses are conducted as objective Bayesian frameworks [15,18].

Similar discussions have been provided in relation to the undercoverage problem of standard confidence intervals of grand mean parameter $\mu$, and various improved methods have been developed [9-11]. Besides, for prediction intervals, rich methods do not yet exist. Currently, the confidence distribution approach of Nagashima et al. [21,23] represents a suitable choice for accurate predictions; in their simulation-based numerical evaluations, they exhibited accurate coverage properties in general. In this regard, development of more rich methods for accurate predictions is needed, and represents an important priority for future work in research synthesis methodology.

## Data Availability Statement

The meta-analysis datasets used in Section 5 are provided in the original papers [4,37-43].

## Acknowledgements

This study was supported by Grant-in-Aid for Scientific Research from the Japan Society for the Promotion of Science (Grant numbers: JP19H04074, JP17K19808).

**Table 1**. Summary of the eight real data examples in Section 5 [†].

| Dataset | Reference | Purpose | Tested treatment (N) | Control treatment (N) | n | Outcome | Type of outcome variable | Effect measure | Overall estimate of $\mu$ (95% CI) | $\hat{\tau}^2_{DL}$ | $I^2$ | p-value of Q-test |
|---|---|---|---|---|---|---|---|---|---|---|---|---|
| CPR data | Hüpfl M et al. [37] | To assess the association of chest-compression-only cardiopulmonary resuscitation with survival in patients with out-of-hospital cardiac arrest | Chest-compression-only cardiopulmonary resuscitation (1500) | Standard cardiopulmonary resuscitation (1531) | 3 | Survival to hospital discharge or 30-day survival | Binary | RR | 1.215 (1.009, 1.464) | 0.00 | 0.0% | 0.6845 |
| Corticosteroids data | Sadeghirad B et al. [38] | To estimate the benefits and harms of using corticosteroids as an adjunct treatment for sore throat | Corticosteroids (439) | Placebo (424) | 5 | Complete resolution of pain | Binary | RR | 2.233 (1.177, 4.235) | 0.33 | 67.9% | 0.0144 |
| QOL data | Gaertner J et al. [39] | To assess the effect of specialist palliative care on quality of life | Specialist palliative care (733) | Standard care (652) | 7 | Quality of life | Continuous | SMD | 0.569 (−0.017, 1.154) | 0.60 | 96.2% | <0.0001 |
| PPI data | Crocker JC et al. [40] | To investigate the impact of patient and public involvement on rates of enrolment and retention in clinical trials | Patient and public involvement (9940) | Non-patient and public involvement (8002) | 8 | Patient enrolment rate | Binary | OR | 1.165 (1.039, 1.306) | 0.00 | 0.0% | 0.4952 |
| SBP data | Riley RD et al. [4] | A hypothetical meta-analysis that assesses the blood pressure reduction effect of an antihypertensive drug | - | - | 10 | Systolic blood pressure (mmHg) | Continuous | SMD | −0.334 (−0.484, −0.184) | 0.03 | 70.5% | 0.0004 |
| DPP-4 data | Salvo S et al. [41] | To quantify the risk of hypoglycaemia associated with the concomitant use of dipeptidyl peptidase-4 inhibitors and sulphonylureas compared with placebo and sulphonylureas | Dipeptidyl peptidase-4 inhibitors and sulphonylureas (4020) | Placebo and sulphonylureas (2526) | 10 | Incidence of hypoglycaemia | Binary | RR | 1.513 (1.219, 1.878) | 0.02 | 19.0% | 0.2680 |
| Breakfast data | Sievert K et al. [42] | To examine the effect of regular breakfast consumption on weight change | Breakfast (213) | Non-breakfast (216) | 10 | Weight change (kg) | Continuous | MD | 0.440 (0.065, 0.816) | 0.13 | 42.9% | 0.0722 |
| Pain data | Häuser W et al. [43] | To determine the efficacy of antidepressants in the treatment of Fibromyalgia syndrome | Antidepressants (1153) | Placebo (1143) | 22 | Pain questionnaire | Continuous | SMD | −0.427 (−0.553, −0.302) | 0.03 | 44.9% | 0.0124 |

[†] CI: confidence interval, RR: risk ratio, SMD: standardized mean difference, OR: odds ratio, MD: mean difference.

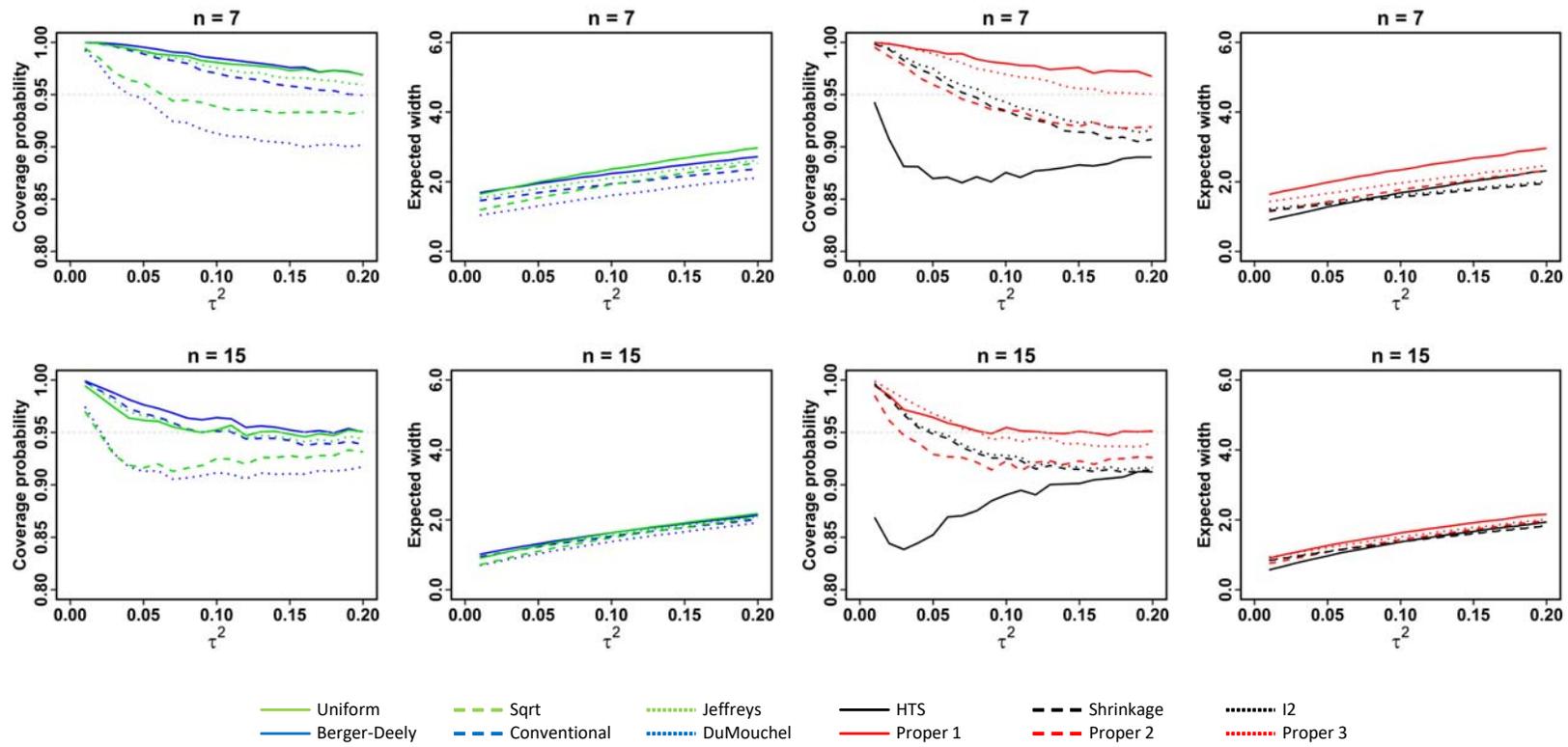

**Figure 1.** Simulation results: coverage probabilities, and expected widths of the 95% prediction intervals. Varying $\tau^2$ on 0.01, 0.02, 0.03,…, 0.20 under fixed $n$ (=7, 15).

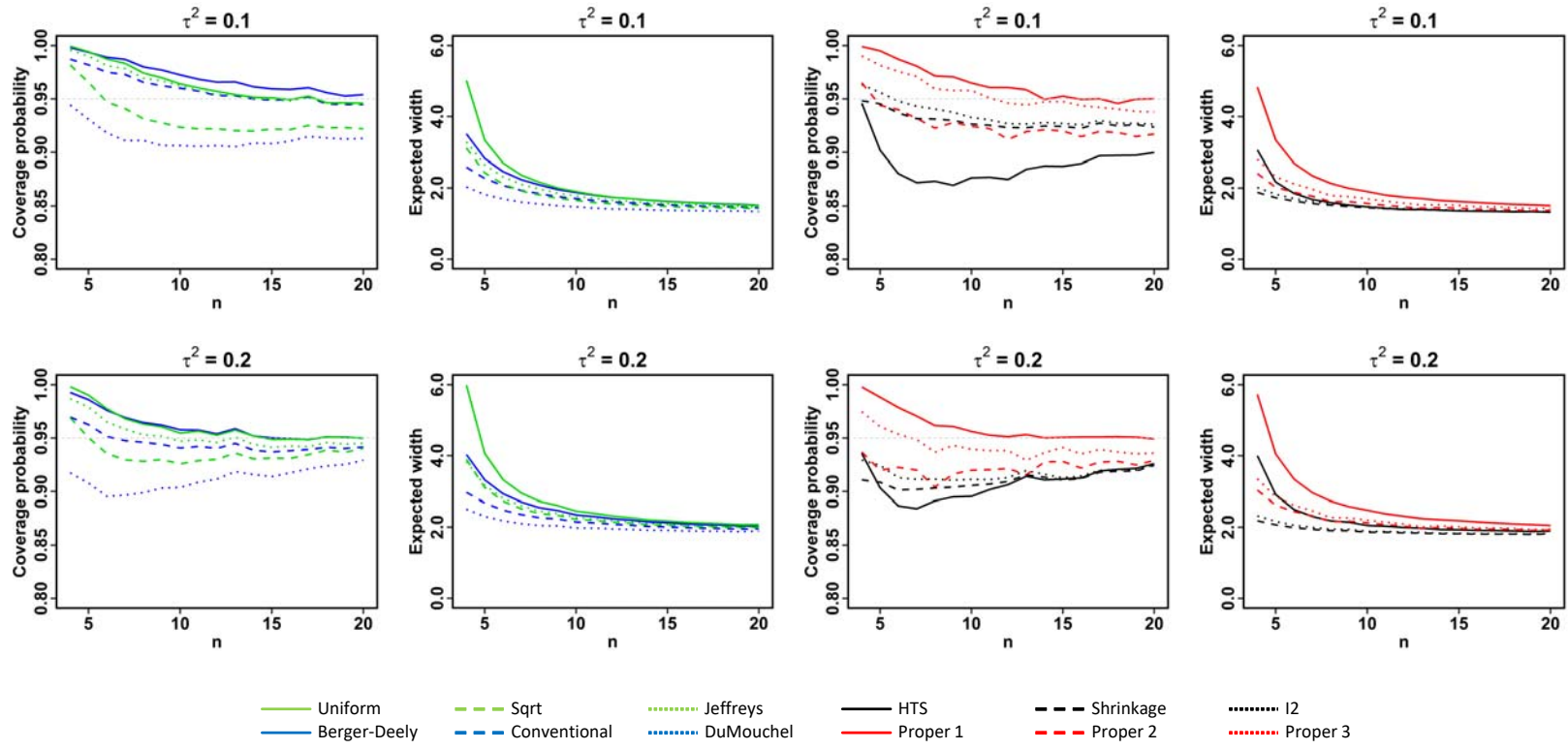

**Figure 2.** Simulation results: coverage probabilities and expected widths of the 95% prediction intervals. Varying *n* on 4, 5, 6,…, 20 under fixed $\tau^2$ (=0.1, 0.2).

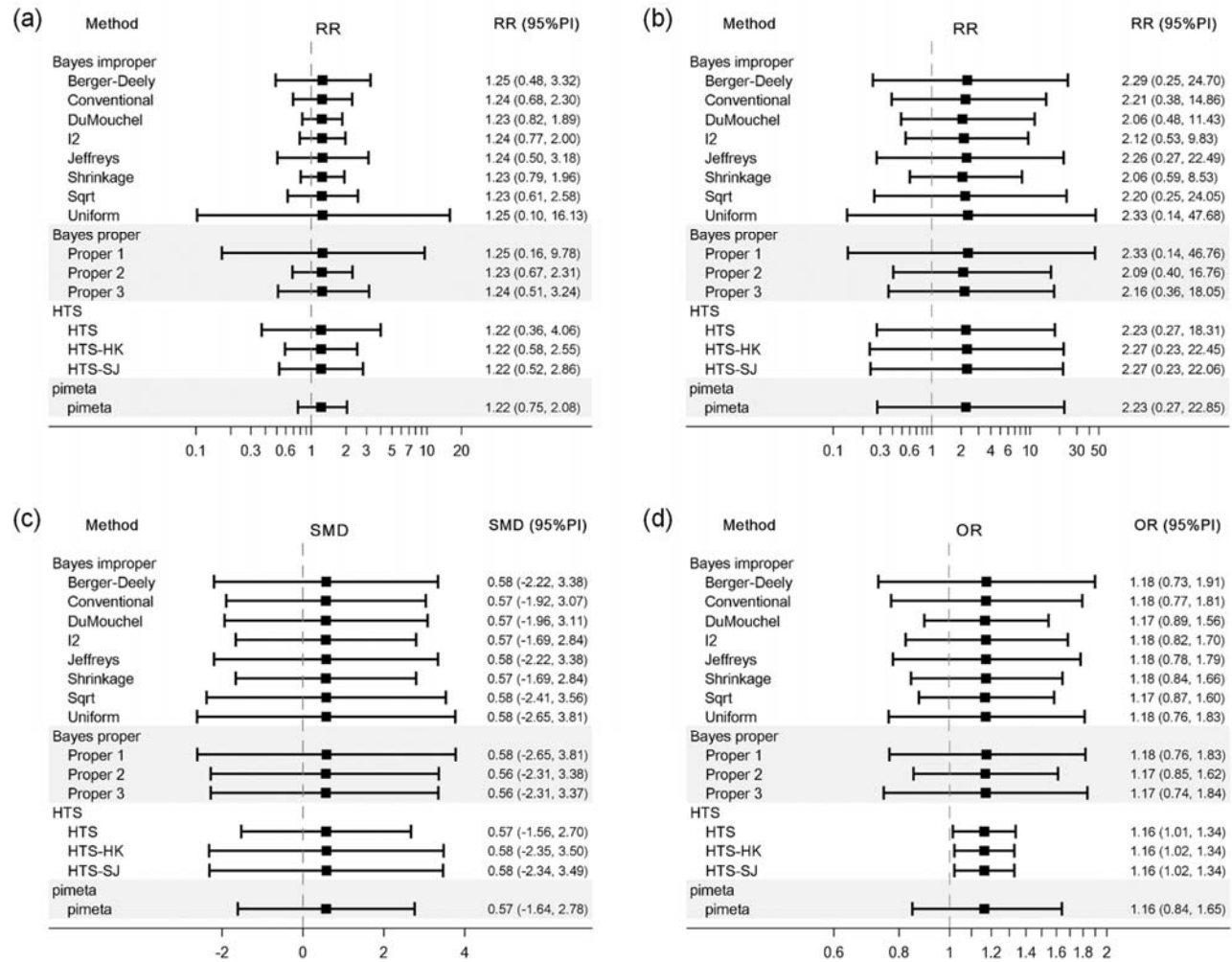

**Figure 3.** The 95% prediction intervals for four illustrative examples: (a) CPR data ($n = 3$), (b) corticosteroids data ($n = 5$), (c) QOL data ($K = 7$), and (d) PPI data ($n = 8$).

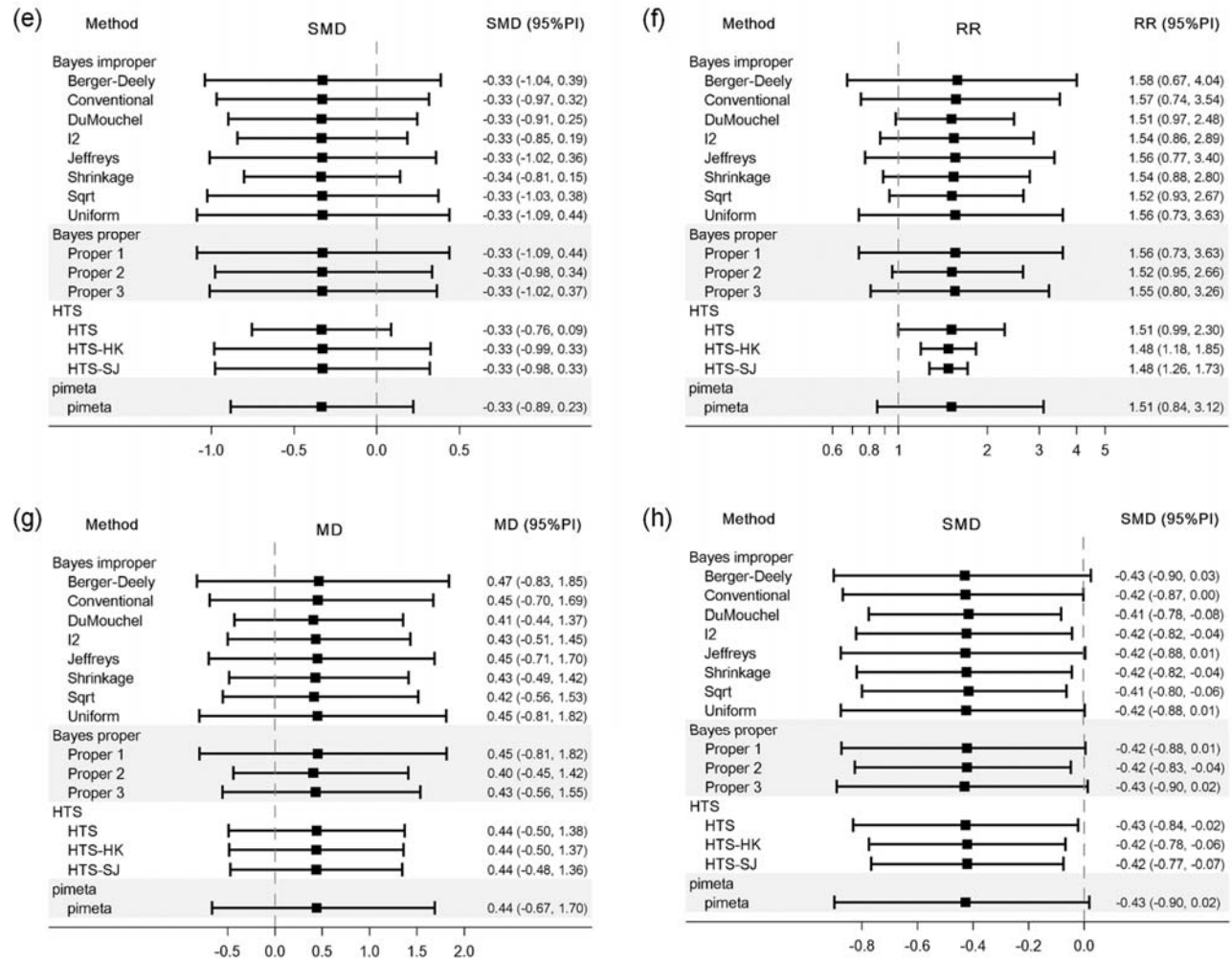

**Figure 4.** The 95% prediction intervals for four illustrative examples: (e) SBP data ($n = 10$), (f) DPP-4 data ($n = 10$), (g) breakfast data ($n = 10$), and (h) pain data ($n = 22$).

# Frequentist performances of Bayesian prediction intervals for random-effects meta-analysis

Hamaguchi, Y.[1,2], Noma, H.[3], Nagashima, K.[4] Yamada, T.[5] and Furukawa, T. A.[6]

[1] *Department of Statistical Science, School of Multidisciplinary Sciences, The Graduate University for Advanced Studies, Tokyo, Japan*

[2] *Diagnostics Department, Asahi Kasei Pharma Corporation, Tokyo, Japan*

[3] *Department of Data Science, The Institute of Statistical Mathematics, Tokyo, Japan*

[4] *Research Center for Medical and Health Data Science, The Institute of Statistical Mathematics, Tokyo, Japan*

[5] *Department of Diabetes and Metabolic Diseases, Graduate School of Medicine, The University of Tokyo, Tokyo, Japan*

[6] *Departments of Health Promotion and Human Behavior, Kyoto University Graduate School of Medicine/School of Public Health, Kyoto, Japan*

## e-Appendix A: Supplementary simulation results in Section 4

In e-Figures 1 and 2, we provide supplementary simulation results for credible intervals that were obtained from the simulation studies described in Section 4. In these figures, we presented empirical coverage probabilities and expected widths of the 95% credible intervals of $\mu$ obtained from the 11 prior distributions. As a reference method, we also present those of the 95% DerSimonian-Laird (**DL**) confidence interval of $\mu$ (DerSimonian and Laird, 1986). Similar simulation-based evidence for Bayesian inferences was also provided by Lambert et al. (2005). The simulation results presented here adopted different prior distributions and were obtained from different simulation settings; therefore, they provide useful numerical evidence for assessing the frequentist coverage performances of Bayesian inference methods in meta-analyses. The overall trends of the coverage properties and expected widths from the 11 priors were similar to those of the prediction intervals (Section 4).

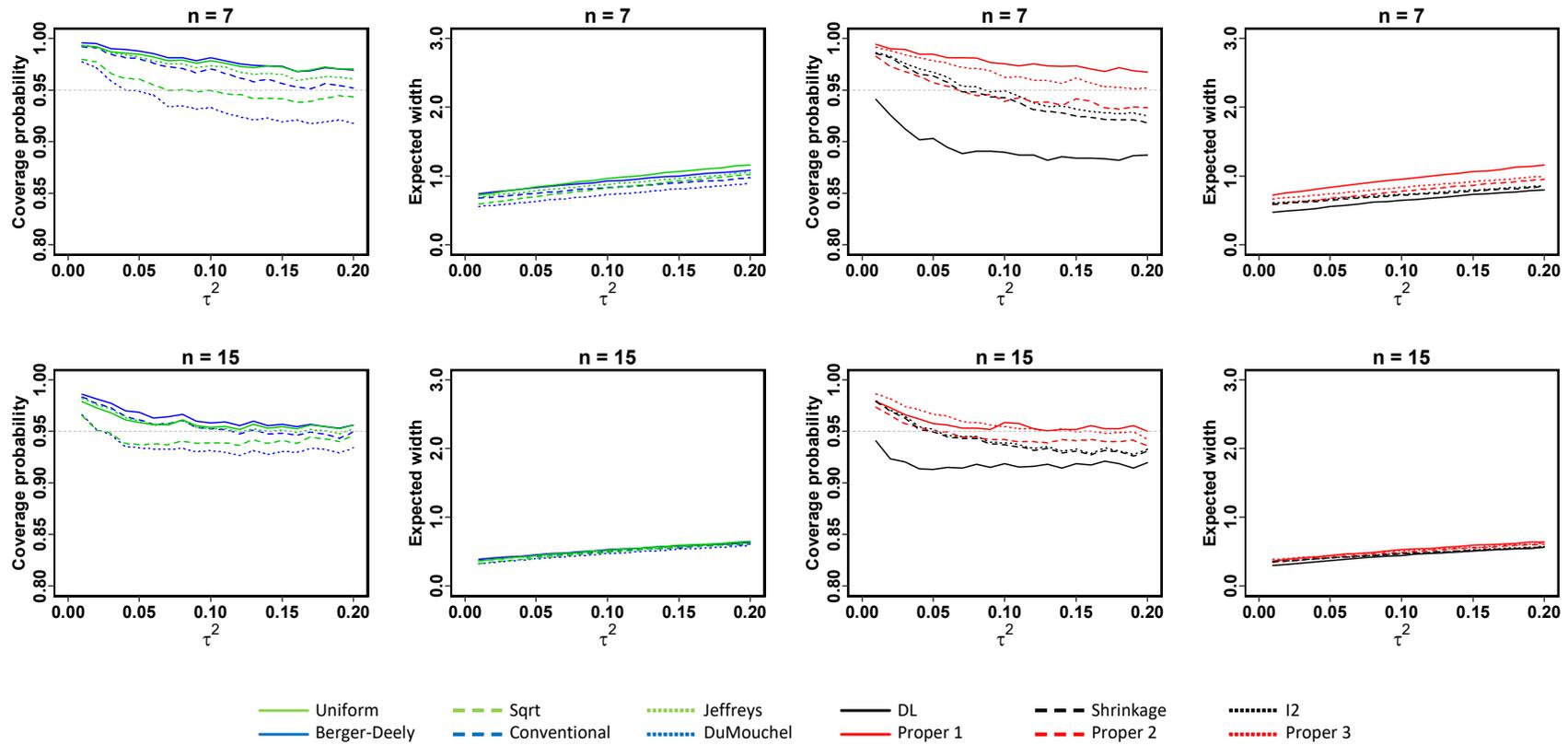

**e-Figure 1.** Simulation results: coverage probabilities and expected widths of the 95% credible intervals of $\mu$. Varying $\tau^2$ for 0.01, 0.02, 0.03,…, 0.20 under fixed $n$ (=7, 15).

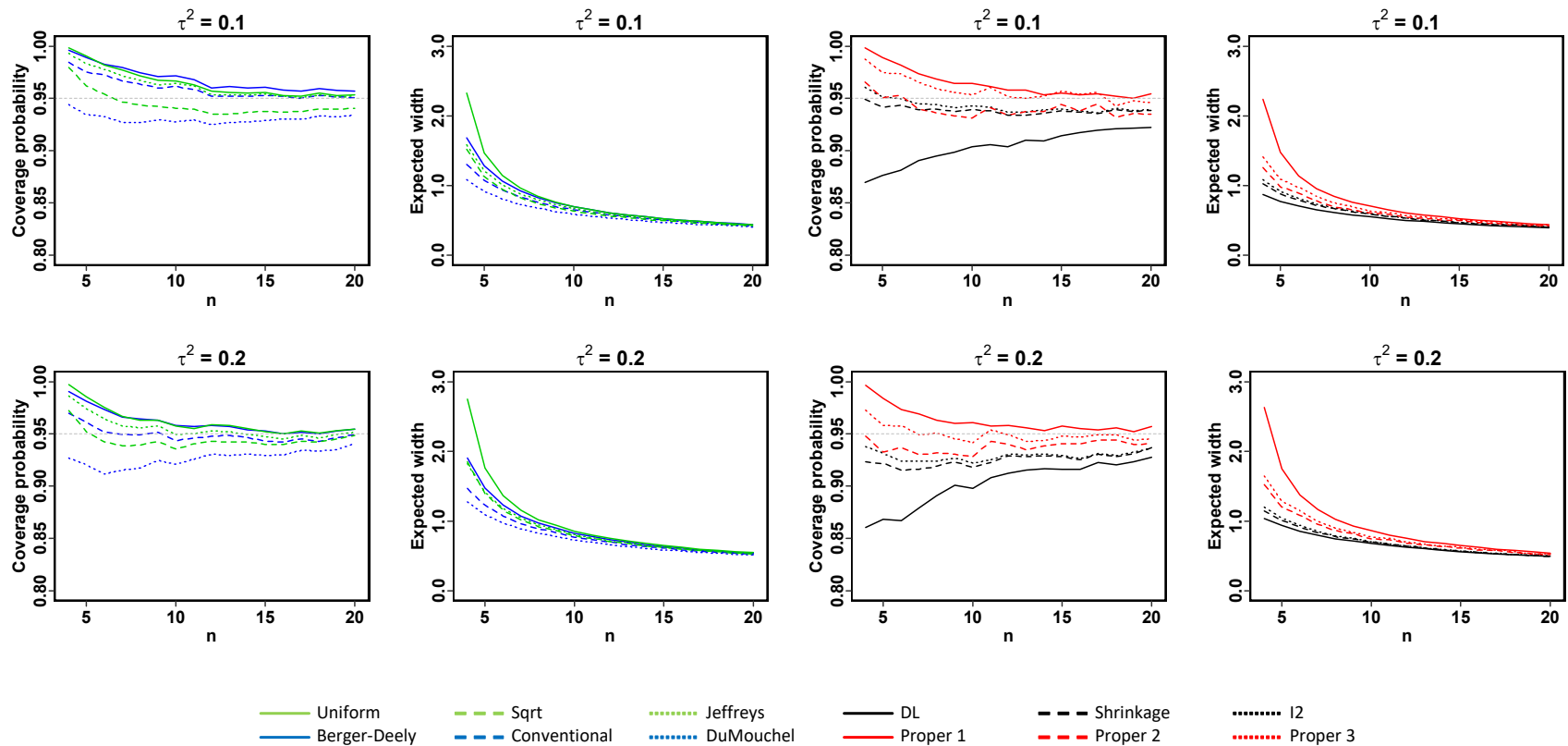

**e-Figure 2.** Simulation results: coverage probabilities and expected widths of the 95% credible intervals of $\mu$. Varying $n$ for 4, 5, 6,…, 20 under fixed $\tau^2$ (=0.1, 0.2).

**e-Appendix B: Credible intervals for the real data analyses in Section 5**

In Section 5, we provide supplementary analysis results for eight real data examples (Crocker et al., 2018; Gaertner et al., 2017; Hauser et al., 2009; Hupfl, Selig and Nagele, 2010; Riley, Higgins and Deeks, 2011; Sadeghirad et al., 2017; Salvo et al., 2016; Sievert et al., 2019); the 95% credible intervals obtained from the 11 priors are presented in e-Figures 3 and 4. As a reference method, we also present the 95% DL confidence interval of $\mu$ (DerSimonian and Laird, 1986). For descriptions of the eight datasets, please see Table 1 in Section 5.

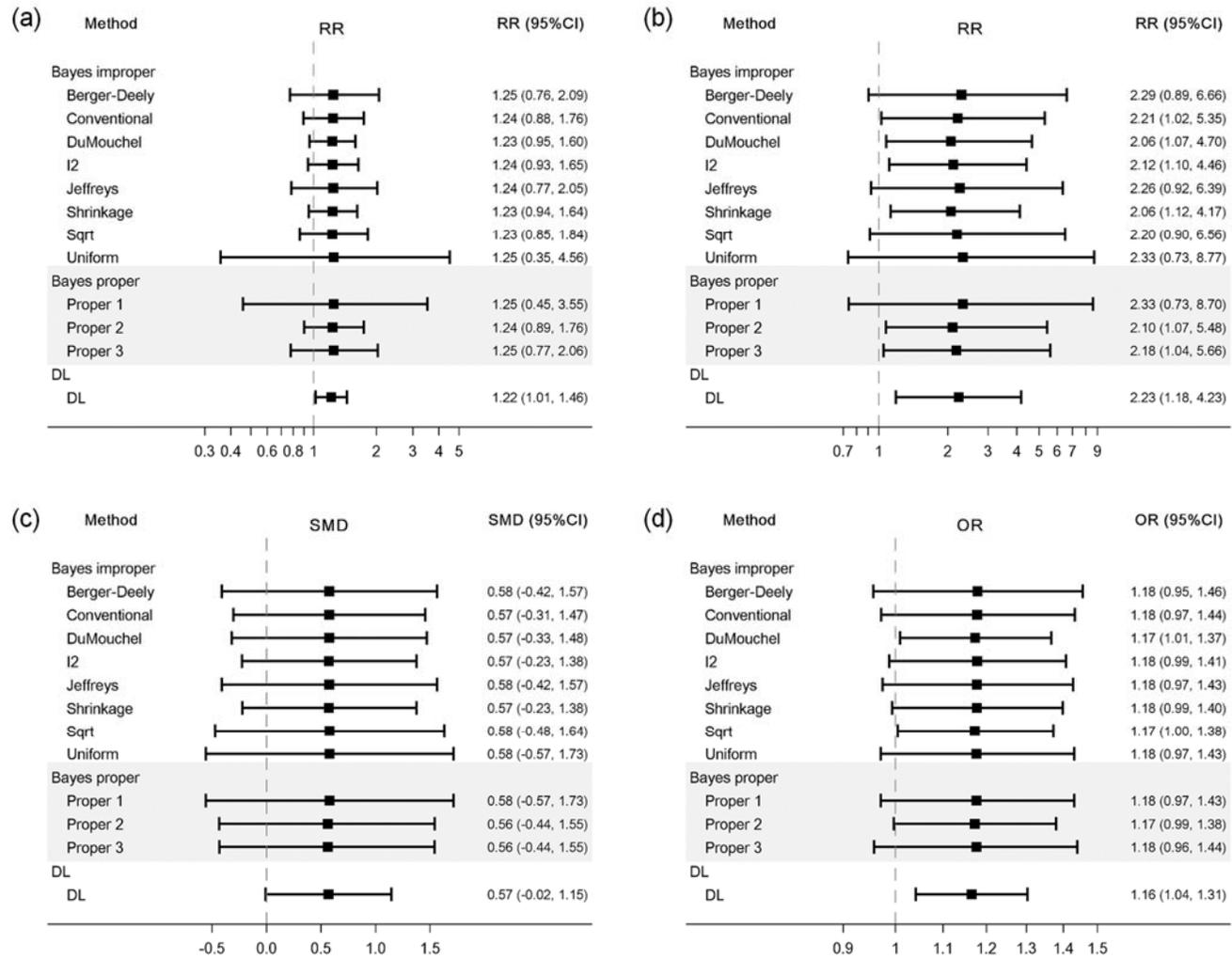

e-Figure 3. The 95% credible intervals of the grand mean parameter: (a) CPR data ($n = 3$), (b) corticosteroids data ($n = 5$), (c) QOL data ($n = 7$), and (d) PPI data ($n = 8$).

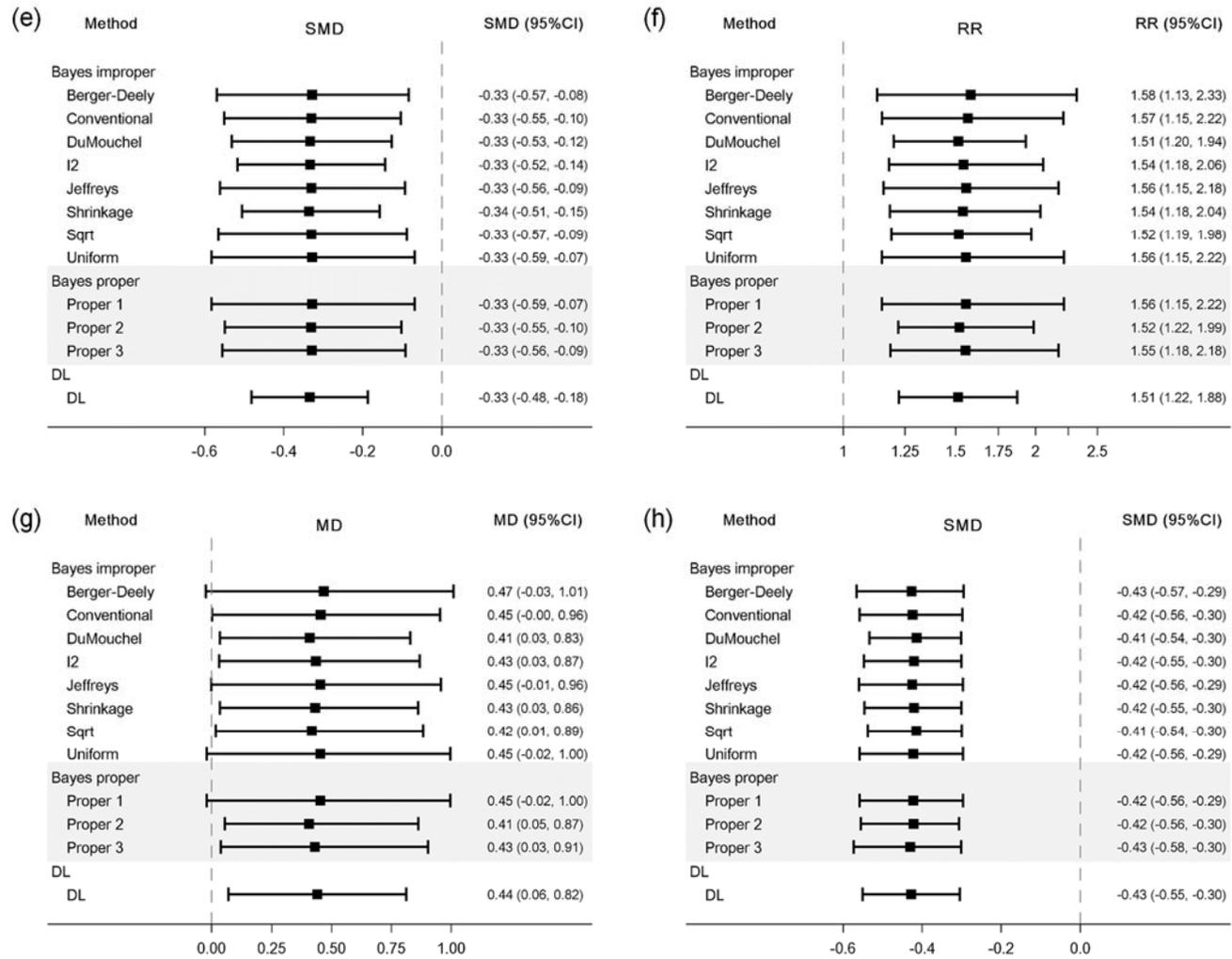

**e-Figure 4.** The 95% credible intervals of the grand mean parameter: (e) SBP data ($n = 10$), (f) DPP-4 data ($n = 10$), (g) breakfast data ($n = 10$), and (h) pain data ($n = 22$).